# Dynamics and Patterning of Screw Dislocations in Two Dimensions

**Robin L. B. Selinger, Brian B. Smith, and Wei-Dong Luo**
Physics Department, Catholic University
Washington, DC 20064
selinger@cua.edu

## ABSTRACT

To understand how dislocations form ordered structures during the deformation of metals, we perform computer simulation studies of the dynamics and patterning of screw dislocations in two dimensions. The simulation is carried out using an idealized atomistic model with anti-plane displacements only; we show that this system is an analog of the two-dimensional XY rotor model. Simulation studies show that under a constant applied shear strain rate, the flow of dislocations spontaneously coalesces to form narrow dislocation-rich channels separated by wide dislocation-free regions, so that the applied strain is localized into slip bands. We argue that this pattern formation represents a phase separation into low/high defect density phases associated with the XY model, and conjecture that thermodynamic forces drive strain localization.

## INTRODUCTION

Much research and indeed much of the work presented in this symposium have been directed at understanding the evolution of dislocation microstructures and the mechanical response of metals. While many insights can be found from mesoscale models, the problem remains essentially unsolved: the elastic-plastic response of metals under arbitrary temperature and deformation history remains impossible to predict from first principles. Many researchers treat dislocation dynamics and patterning as a complex non-equilibrium reaction-diffusion process and focus on enumerating the list of potential dislocation reactions, sometimes relying on atomistic simulation to extract rules and parameters for use in mesoscale simulations.

Here we take a rather different approach, and study the dynamics and patterning of screw dislocations under a constant applied shear strain rate in two dimensions, using a simplified atomistic simulation. Although the 2-d geometry neglects dislocation entanglement and a host of other 3-d phenomena, the simulation shows formation of slip bands with a spacing that depends on the shear strain rate. We map the atomistic model onto a statistical physics model, the XY rotor model, to gain insight into the thermodynamic forces that drive the localization of strain. By understanding the mechanisms for strain localization in this highly idealized 2-d system, we hope to make progress toward a solution of the more complex 3-d problem.

## MODEL

Consider a simple cubic lattice of particles, where each vertical column moves as a unit and can be displaced only along the column axis, as shown in Fig. 1(a). This system can contain straight parallel screw dislocations, as shown in Fig. 1(b), but cannot contain edge dislocations, as there are no in-plane displacements. The deformation can be represented as a scalar field $z(x,y)$.

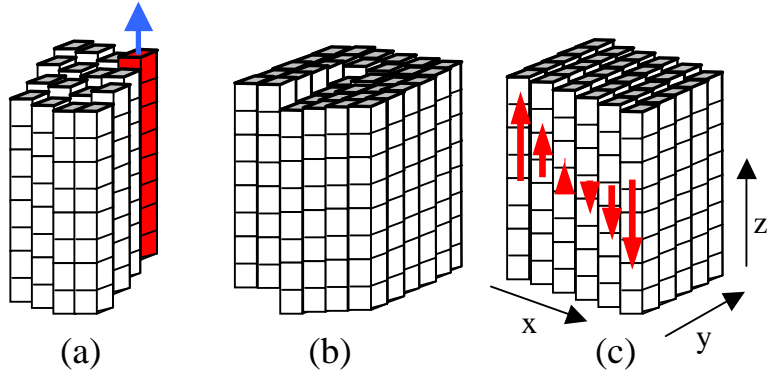

**Fig. 1** *(a) Displacement $z_i$ of each column is a dynamic variable. (b) Straight screw dislocations are allowed in this restricted geometry, but no edge dislocations. (c) Driving boundary conditions maintain constant shear strain rate.*

Neglecting any surface effects, the Hamiltonian (per unit length) of this system is:

$$H = -K \sum_{<i,j>} \cos[2\pi(z_i - z_j)/a_o] + \tfrac{1}{2} \sum_i m \left(\frac{dz_i}{dt}\right)^2 \qquad (1)$$

Here $z_i$ is the out-of-plane displacement of column $i$, $K$ is an energy per unit length (closely related to the unstable stacking energy), $a_o$ is the lattice spacing, $m$ is the mass per unit length of each column, and the first sum is over nearest neighbor columns $<i,j>$. The cosine term represents the energy per unit length associated with each pair of neighboring columns as they slide in and out of registry with one another, with an energy barrier of $2K$ to cross from one local energy minimum to the next. The second term represents the kinetic energy per unit length associated with the motion of the columns.

Deterministic equations of motion are derived from the Hamiltonian, just as in classical molecular dynamics (MD) simulations, and are integrated forward in time with a finite time step. Typical system size studied is $1{,}000 \times 200$ and simulations are run up to $2 \times 10^6$ discrete time steps, for total duration of $10^5\, t_o$. Here $t_o = a_o\, (m/K)^{1/2} \approx 10^{-12}$ sec is the natural time unit.

We use this model to study the homogeneous nucleation, motion, and pattern formation of screw dislocations under a constant shear strain rate $\dot{\varepsilon}$ at temperature $T$. In the initial conditions, each column starts with zero displacement and a small randomized z-velocity appropriate for the chosen temperature. A velocity gradient is then superimposed along the x direction as shown in Fig. 1(c). Periodic boundary conditions are applied with a z-offset across the boundary in the x direction. The offset increases linearly with time, and this "driving" boundary condition maintains the total shear strain rate $\dot{\varepsilon}$ at a fixed value. Regular periodic boundary conditions are used in the y-direction. The driving boundary conditions do work on the system, so a thermostat algorithm is used to maintain constant T.

To make the model more realistic, the cosine term in Eqn. 1 could be replaced with an appropriate Embedded Atom Method potential, and a slice of a suitably oriented FCC lattice could replace the square lattice. However the highly idealized version of the model is of particular interest because of its close analogy to the XY rotor model. The 2-d XY rotor model consists of a lattice of spins as shown in Fig. 2. Each spin lies in the x-y plane and has an angular orientation $\theta_i$ with respect to the x axis, with $0 < \theta_i < 2\pi$. The Hamiltonian is:

$$H = \sum_{<i,j>} -J\, \cos(\theta_i - \theta_j) + \sum_i \tfrac{1}{2} I \left(\frac{d\theta_i}{dt}\right)^2 \qquad (2)$$

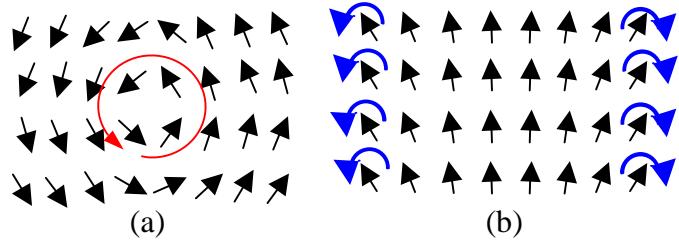

*Fig. 2*: *(a) Positive vortex, analog of a positive screw dislocation. (b) The XY rotor model under "twisting" boundary conditions, analogous to the applied shear strain shown in Fig. 1(c).*

Here *J* is the (positive) exchange coefficient, and *I* is the moment of inertia of each spin, treated as a rigid rotor.

The mapping between the two models involves the transformation, $\theta_i = 2\pi\, z_i / a_o$. A screw dislocation in the atomistic model maps to a vortex in the XY rotor model, as shown in Fig. 2(a). Applied shear strain in the crystal is analogous to applied twist in the XY rotor model, shown in Fig. 2(b). The sign and "charge" of a vortex can be calculated by making the equivalent of a Burgers circuit around the defect core. If the result is $+/-2\pi$, the vortex is positive/negative, respectively. If the result is 0, there is no defect. See Chaiken and Lubensky [1] for a detailed discussion of topological defects in the XY model.

The XY model in 2-d lacks true long-range order at low temperature due to thermal fluctuations. The system has only a small density of thermally activated vortices, and these are bound in closely-spaced pairs. At a critical temperature of approximately $k_B T/J = 0.89$, the system undergoes a phase transition, known as the Kosterlitz-Thouless (K-T) transition [2]. This is a second-order melting transition that occurs via cooperative homogeneous nucleation and unbinding of vortex dipoles. The high T phase has a high density of unbound vortices.

The addition of driving boundary conditions changes the phase behavior of the XY rotor model, because stress drives the nucleation and unbinding of vortex pairs even at low temperature. Reduction of the effective transition temperature due to stress was predicted theoretically by Khantha et al [3] in their theory of the ductile-brittle transition. We find that the critical temperature for defect nucleation drops linearly with increasing stress [4] in agreement with their predictions.

**RESULTS**

We carry out simulations at T=0.2, well below the K-T transition, with the system initially vortex-free. The shear strain rate is applied, and at a critical value of the stress, dislocation pairs nucleate and unbind in roughly constant density, as shown in Fig. 3(a). Positive and negative screw dislocations are driven in opposite directions by the stress and undergo more or less constant motion, with frequent nucleation/annihilation, and the structure is in continuous evolution. Dislocation flow then spontaneously localizes into a nearly periodic array of narrow channels separated by wide defect-free regions, as shown in Fig. 3(b). The initial spacing of the channels depends on the strain rate; higher strain rate produces more closely spaced channels. Once coalesced into channels, dislocations continue to flow in opposite directions with frequent nucleation/annihilation events. As there is no pinning mechanism, there are no immobile dislocations. Fig. 4 displays the z(x,y) displacement field for a larger system, showing that the dislocation-rich channels represent localization of the shear strain into a series of nearly periodic slip bands.

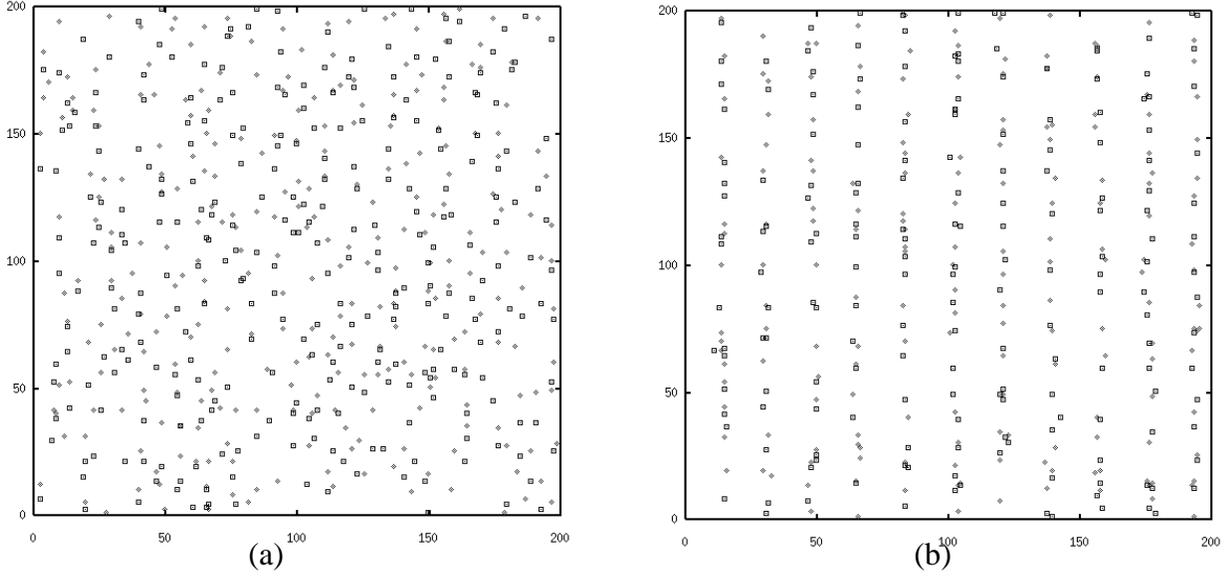

***Fig. 3***: *Simulation with T=0.2 and $\dot{\varepsilon} = 3.2 \times 10^{-3}$, showing topological defects in the **x-y** plane. (Note that the strain rate is defined per natural time unit $t_o$.) Screw dislocations of opposite signs are displayed as ◆ and ■. (a) Initial density is uniform. (b) Dislocation channels form. Defects of opposite sign travel in opposite directions along each channel.*

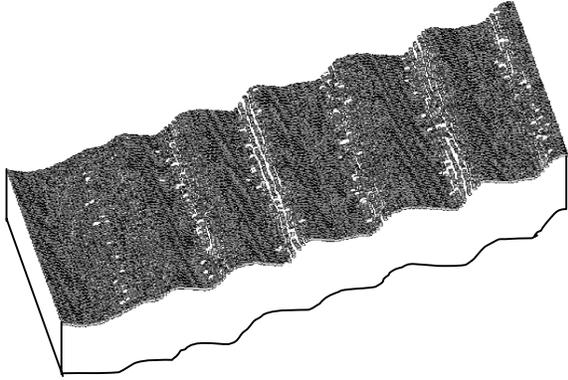

***Fig. 4:*** *Three-dimensional displacement field $z(x,y)$ for a system of size $1000 \times 200$. Dislocation channeling produces localization of the applied shear strain into a series of slip bands.*

The channel pattern coarsens, as channels diffuse laterally and occasionally merge. Channel number drops as $t^{-\alpha}$ where $\alpha < 0.5$ depends on the strain rate. The overall dislocation density remains roughly constant over time; channel width increases as the pattern coarsens. Equilibrium dislocation density scales with strain rate with an exponent of 0.75 ±0.05, as shown in Fig. 5. Since scaling exponents generally depend sensitively on dimensionality, this exponent cannot be usefully compared with experimental results for three-dimensional materials.

**DISCUSSION**

The close analogy between anti-plane deformation of a crystal and the XY rotor model provides a useful theoretical framework for understanding the coalescence of dislocation flow into channels and the localization of strain. In the absence of driving boundary conditions, the XY model has no unbound vortices at low temperature. The imposition of driving boundary conditions imposes a non-zero density of free topological defects, because $\dot{\varepsilon} = \rho b v$, where $\rho$ is the total defect density ($\rho_- + \rho_+$), $v$ is the average speed of defect motion, and b is Burgers vector. In principle the induced defect density could remain uniform, but the system is not stable at arbitrary defect density and divides into a nearly defect-free phase and a defect-rich phase. We conjecture that the system lowers its free energy by phase separating, that is, two-phase coexistence is lower in free energy than the homogeneous phase. The defect-free phase is stiff with a well-defined elastic modulus, while the defect-rich phase yields with a viscous response. The coarsening process we observe can thus be interpreted as a simple phase separation and coarsening of a system in two-phase coexistence. The driving force for later channel coalescence is surface tension associated with the boundary between the two phases. In future work we hope to complete an analytical calculation of the free energy density associated with the two phases and derive analytical results for the thermodynamic forces driving phase separation. Such a result would greatly clarify the role of equilibrium thermodynamics in the pattern formation process, a matter of some controversy since the system is dissipative and driven by external forces, and is not in thermal equilibrium in the usual sense.

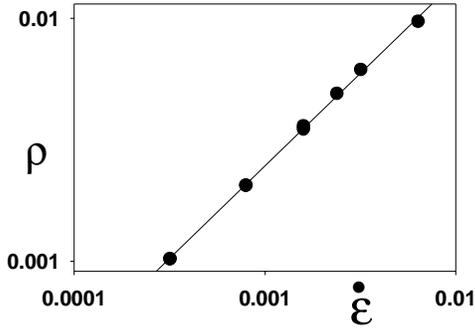

*Fig. 5: Dislocation density $\rho = \rho_+ + \rho_-$ scales with strain rate as $\rho \propto (\dot{\varepsilon})^{3/4}$.*

Holt [5] argued that a gas of +/- screw dislocations in 2-d is unstable to the formation of cell structure through spinodal decomposition. In the absence of an applied stress or strain, the defects would annihilate, but Holt rather unphysically assumed fixed density. In our simulations, we find that patterning occurs only in the presence of an applied strain rate, a more physical way to impose non-zero defect density. Further analysis of the thermodynamics of slip band formation and coarsening may also explain the ubiquitous scaling behavior seen by Hughes et al [6] in dislocation microstructures.

## APPLICATIONS

Besides providing a novel theoretical viewpoint on dislocation patterning, the two-dimensional deformation model presented here can be used to investigate a variety of phenomena related to the elastic-plastic response of crystalline solids. A crack may be introduced into the model by allowing the coefficient $K$ in Eqn. [1] (equivalently $J$ in Eqn. [2]) to vary as a function of position, setting it equal to zero along part of a row of nearest-neighbor bonds. Stress intensification occurs at the crack tips in the normal way and heterogeneous screw dislocation nucleation is observed under shear, which loads the crack in mode III . We introduce a bond-breaking rule that sets $K$ on a bond to zero when the local stress exceeds a threshold value. Preliminary studies indicate that with this modification, the system displays a brittle-ductile transition with temperature and strain rate. This model will allow direct calculation of fracture toughness as a function of temperature and strain rate in two dimensions, and should provide a useful test for comparison with the predictions of the theory by Khantha et al [3].

This model may also have use in the study of both size effects and strain-gradient effects in plasticity. Even in a simple geometry like the one explored here, one can expect size effects if the sample is comparable in size to the equilibrium channel spacing. Strain gradients can be introduced by changing from simple shear to more complex geometries. For example, we have started initial studies on the shear of a thick-walled cylindrical shell between its inner and outer surfaces, and of simple torsion of a beam.

As discussed above, the model can also be altered to include realistic potentials and lattice structure and to include both in-plane and out-of plane displacements. At that stage the model becomes essentially a molecular dynamics simulation, but with a particular 2-d geometry that could display dislocation patterning at accessible time and length scales. Even in its simplest form, the present model has advantages over discrete dislocation dynamics techniques. Pair nucleation, defect motion, and annihilation occur spontaneously through the dynamics of the model with no need for arbitrary rules or extra parameters. Since it is an atomistic model, though a simplified one, dislocations interact through a dynamic strain field. They have appropriate effective mass, velocity-dependent damping, and relativistic effects.

## CONCLUSIONS

We have carried out simulation studies of screw dislocation dynamics and patterning in two dimensions using a simplified atomistic model. The simulations show the formation of distinct dislocation channels, or slip bands, with accompanying localization of applied shear strain. The initial spacing between slip bands drops with increasing strain rate. Total dislocation density appears to scale with strain rate with a characteristic exponent of approximately ¾.

We use the close analogy between anti-plane shear deformation of a crystal and the XY rotor model to gain new insight into the mechanisms driving dislocation patterning and strain localization in two dimensions. We argue that the mechanism driving this pattern formation is a phase separation into two phases characterized by low and high defect densities. The simulation model shows promise for further investigation of elastic-plastic response of solids, including size effects and strain gradient effects in plasticity and the brittle-ductile transition.

## ACKNOWLEDGMENTS

This work was supported by The National Science Foundation NSF-DMR-9702234.